\documentclass{aa}
\usepackage{psfig}

\begin{document}

\author{A.~Tokovinin\inst{1}, S.~Thomas\inst{1}, M.~Sterzik\inst{2},
S.~Udry\inst{3}
}

\institute{
Cerro Tololo Inter-American Observatory, Casilla~603, La Serena,
Chile %
\and
European Southern Observatory, Casilla~19001, Santiago~19, Chile.
\email{msterzik@eso.org}
\and
Observatoire de Gen\`eve, CH-1290 Sauverny, Switzerland
\email{Stephane.Udry@obs.unige.ch}
}

\offprints{A. Tokovinin, \email{atokovinin@ctio.noao.edu} }
\authorrunning{Tokovinin et al.}
\date{Received  / Accepted }

\title{Tertiary    companions   to   close    spectroscopic   binaries
\thanks{Based on observing runs  74.C-0074 and 75.C-0112 at the Very
Large  Telescope of the  European Southern  Observatory at  Paranal,
Chile.
Tables 2 and 5  are only available in electronic form
at the CDS via anonymous ftp to cdsarc.u-strasbg.fr (130.79.128.5)
or via http://cdsweb.u-strasbg.fr/cgi-bin/qcat?J/A+A/
} }

\titlerunning{Tertiary companions to SB}

\abstract{We have surveyed a sample of 165 solar-type spectroscopic
binaries (SB) with periods from 1 to 30 days for higher-order
multiplicity. A subsample of 62 targets were observed with the NACO
adaptive optics system and 13 new physical tertiary companions were
detected.  An additional 12 new wide companions (5 still tentative)
were found using the 2MASS all-sky survey.  The binaries belong to 161
stellar systems; of these 64 are triple, 11 quadruple and 7 quintuple.
After correction for incompleteness, the fraction of SBs with
additional companions is found to be 63\% $\pm$ 5\%.  We find that
this fraction is a strong function of the SB period $P$, reaching 96\%
for $P<3^d$ and dropping to 34\% for $P>12^d$. Period distributions
of SBs with and without tertiaries are significantly different, but
their mass ratio distributions are identical. The statistical data on
the multiplicity of close SBs presented in this paper indicates that
the periods and mass ratios of SBs were established very early, but
the periods of SB systems with triples were further shortened by
angular momentum exchange with companions.
  }

\maketitle

\section{Introduction}
\label{sec:intro}

The formation of binary stars remains a  subject of active
research and debate (Zinnecker \& Mathieu \cite{IAU200}).  Close
binaries are particularly difficult to explain because at orbital
periods of a few days, the separations are less than the size of the
individual components during  the protostellar phase.  Fission of
a rapidly rotating pre-main sequence stellar configuration appears
unlikely according to the most recent theoretical results (Tohline
\cite{Toh02}).  Thus, some mechanism(s) of orbit shrinkage must be
present in order to extract the angular momentum from a proto-binary.
This momentum could be deposited in wider stellar companions -- a
hypothesis we test in this paper.

Stars form in groups and interact dynamically. Detailed studies of
dynamical decay of small unstable clusters, e.g. by Sterzik \& Durisen
(\cite{SD98}), were successful in explaining such properties as
multiplicity rate and mass ratio distribution.  But dynamical
interactions within star clusters alone cannot sufficiently broaden an
initially narrow separation distribution (Kroupa \& Burkert
\cite{KB01}).  Additional dissipative processes and even higher
stellar densities are required (Sterzik, Durisen \& Zinnecker
\cite{SDZ04}).

Larson (\cite{Larson02}) argues that tidal torques excited by stellar
companions in an accretion disk are the most likely and efficient
mechanism to extract angular momentum from accreting stars. In the
absence of angular momentum transport, accreting matter
accumulates in a disk or torus that fragments rapidly, creating the
companion.  Indeed, hydrodynamical simulations of Bate et al.
(\cite{Bate02}) show how such companions are formed and how accreting
binaries become tighter by interacting with their distant companions
or with fly-by members of  a  nascent cluster.  Reipurth
(\cite{R2000}) gives convincing evidence that strong accretion and jet
activity is actually observed in embedded young multiple systems.  He
relates periodic knots in Herbig-Haro jets to periastron passages in
inner (unresolved) binaries and interprets  the
decreasing separation between the knots as a fossil record of orbit
shrinkage.  However, the triple systems examined by Reipurth and Bate
et al.  are relatively wide systems ($ \sim 10$ A.U.), and cannot
account for the characteristics found in short period
pre-main-sequence (PMS) spectroscopic binaries (SB).  Several PMS
binaries with periods shorter than 2 days are known (Melo et al.
\cite{Melo01}), and the SB fraction among PMS stars may be as high as
among field stars.  Many PMS SBs are actually members of higher-order
multiple systems (Sterzik et al.  \cite{SMTB05}).  A comprehensive
star formation theory must be able to explain these observational
facts.

The discovery of massive extrasolar planets with predominantly short
orbital periods revived the interest in orbital decay via interactions
within disks. Planet migration in a disk is now a generally accepted
theory.  A distant (stellar) companion enhances planet growth and
migration (Zucker \& Mazeh \cite{ZM02}; Udry et al. \cite{Udry03};
Eggenberger et al.  \cite{Egg04}), possibly in an analogous way 
to multiple-star formation as envisioned by Larson (\cite{Larson02})
and others.

Present-day parameters of close
binaries are not identical to their parameters at birth (i.e. when
their masses were assembled) because of subsequent evolution. One such
evolutionary mechanism is called KCTF -- Kozai cycles with tidal
friction (Eggleton \& Kisseleva-Eggleton \cite{KCTF}).  A distant
companion in a stable (hierarchical) triple system causes periodic
modulation of the inner-binary eccentricity by Kozai cycles.  These
cycles are periodic, but only as long as the inner system does not
interact tidally at periastron.  In this case the Kozai cycles
are gradually modified, the inner system gets ``locked'' in a
high-eccentricity state and its orbit then slowly decays to a circular
SB with a period of few days (Kisseleva et al.  \cite{Kisseleva98}).
The period distribution of solar-type close binaries within
higher-order multiple systems seems to match this scenario, showing a
sharp drop in the number of systems with $P>7^d$ (Tokovinin \& Smekhov
\cite{TS02}).

A competing evolutionary effect is the disruption of multiple systems
by dynamical instabilities or by perturbations from other stars
passing nearby.  Thus, SBs formed within higher order multiples can
also lose tertiary companions (TCs) at later times.  Interactions with
field stars can be neglected in the present study because  they
disrupt only very wide tertiaries with separations above 0.1~pc (see
Close et al. \cite{Close90} for a review of observational data and
theory).  Much closer tertiaries with periods as short as $\sim$ 300~yr
could be ``ionizied'' during early evolutionary stages in  very
young clusters, explaining the difference in wide-binary frequency
between young T~Tau associations and the field (Kroupa
\cite{Kroupa01}).

Observers have noticed the high multiplicity of close binaries for
some time.  Mayor \& Mazeh (\cite{MM87}) looked for orbital precession
(presumably caused by TCs) in a sample of 25 solar-type SBs and found
that some 25\% of those binaries are triple.  Isobe et al.
(\cite{Isobe}) searched for visual companions to SBs by means of
speckle interferometry.  Tokovinin (\cite{MSC}, MSC) showed that 43\%
of nearby solar-type stars with periods under 10 days have known
tertiaries.  All five such systems in the Duquennoy \& Mayor
(\cite{DM91}, DM91) G-dwarf sample are triple.

Are {\it all} close binaries triple?  Is the angular
momentum of the close binary system always extracted by a tertiary
component (via the KCTF mechanism or during accretion)?  In the present
paper, we attempt an observational  investigation of the
question and prove that not all close SBs are triple.

The paper is organized as follows.  We describe our sample of close
solar-type binaries in Sect.~\ref{sec:sample}.  New components were
discovered with adaptive optics (Sect.~\ref{sec:AO}) and 
the 2MASS sky survey (Sect.~\ref{sec:2MASS}).  We estimate the
massses and periods of all tertiaries in Sect.~\ref{sec:par} and
describe the detection limits of new and existing techniques in
Sect.~\ref{sec:det}.    We study the
statistics of tertiary components in Sect.~\ref{sec:stat}.
Section~\ref{sec:sum} summarizes the results, and the
implications for close-binary formation theories are discussed in
Sect.~\ref{sec:disc}.

\section{The sample}
\label{sec:sample}

\begin{figure}[ht]
\centerline{\psfig{figure=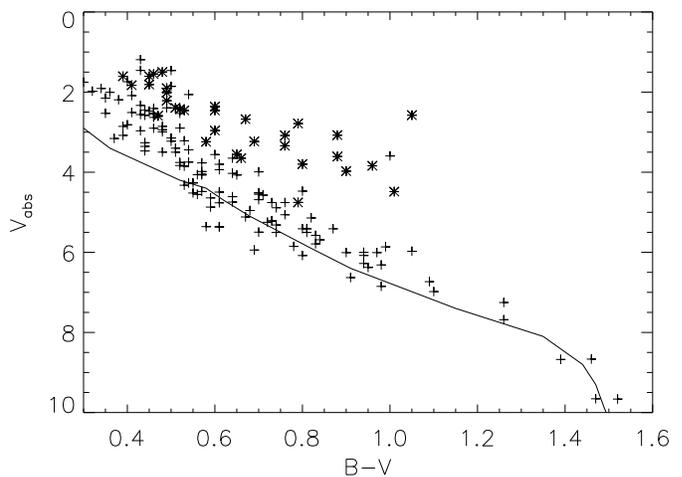,width=10cm} }
\caption{Absolute $V$-magnitudes of the  SB targets in the main
  sample versus their $B-V$ colors. Asterisks mark the SBs with
  evolved components,  the  solid line shows the Main
  Sequence from Lang (1992).
\label{fig:Vabs} }
\end{figure}

We define a sample of spectroscopic binaries (SBs) where the chance of
discovering a TC is maximized. Ideally, the systems should be composed
of low-mass dwarf stars because such stars are not too bright,
increasing the possibility of detection of close visual companions,
and have sharp lines making it easier to detect  spectroscopic
tertiaries.  Furthermore, the stars should be nearby, leading to larger
separation of tertiaries. We also want to probe the frequency of TCs to
SBs with periods below and above the KCTF limit of $P \sim 7^d$, and
hence consider the SBs with periods up to $30^d$.

The main sample was constructed by selecting solar-type SBs from the
catalog of Batten et al.  (\cite{SB8}). No condition on the luminosity
class was imposed:  short orbital periods guarantee
small stellar radii, hence evolved components (giants) will be
excluded. It is known that luminosity classes assigned by observers
are often contradictory, hence we tried to ignore them. It turned out
later that some objects are in fact moderately evolved
(Fig.~\ref{fig:Vabs}).

The sample was complemented by new SBs from recent surveys of nearby
stars (Latham et al.  \cite{L02}; Goldberg et al.  \cite{G02};
Halbwachs et al.  \cite{H03} and other).  New unpublished SBs from the
CORALIE ongoing survey were also added (Table~\ref{tab:COR}).  A total
of 200 SBs  were thus selected.  However, the main sample 
was then restricted to stars with Hipparcos parallaxes larger than
10~mas (i.e.  within 100~pc). This sample contains 165 SBs with
periods less than 30 days belonging to 161 stellar systems 
(including four multiple systems with two short-period SBs each). Each
system is identified by its Hipparcos number (when several components
have distinct HIP numbers, only one is retained).  The main data
on the sample -- HIP and HD numbers, coordinates, proper motions,
parallaxes, magnitudes and spectral types -- are given in Table~2,
available electronically.  Several targets belong to the Hyades
cluster.

\begin{table}
\caption{Preliminary parameters of CORALIE binaries.
The columns contain: (1) HIP number, (2) HD number, (3) orbital period,
(4) semi-amplitude of the primary and (5) Notes.
}
\label{tab:COR}
\begin{tabular}{c c c c l }
\hline
\hline
HIP & HD & $P_1$,  & $K_1$, & Note \\
    &  & d      & km/s   &  \\
\hline
 2790  & 3277   & 15   &   5? &   Uncertain SB  \\
 2848  & 3359   & 22    & 22.4  & B-comp. to HIP 2888\\ 
19248  & 26354  & 2.5   &  56   & $P_3>3000$ d \\
38179  & 64184  &17.8  &  12.7 & \\ 
45957  & 81044  &15  &   8 &  \\ 
49161  & 87007  &30    & 13.4   & $P_3$ suspected \\
53217  & 94340  &6.8   & 20?  & Susp. $P_3 = 1200$ d \\
56960  & 101472 &4.4   & 15.7  & \\
73269  & 132173 &14  & 31.5  & \\
85675  & 158577 &9.69  & 34.5  & \\ 
94863  & 180445 &2.5   & 47  & SB2 \\
97030  & 186160 &10.7  & 18  & \\
107779 & 207450 & 4.9  &  15.8 & \\
114703 & 219175 & 7.1  &  22.9 & CPM HIP 114702 \\
116429 & 221818 &11.6  &  15 & \\
\hline
\end{tabular}
\end{table}

\setcounter{table}{2}

The masses of the primary components of the SBs are either known from
orbital solutions (e.g. eclipsing pairs) or estimated from the $B-V$
colors and spectral types using standard relations for the Main
Sequence (MS) from Lang (\cite{Lang}).  Despite the brightness of our
targets, the task of assembling  the basic data turned out to be
non-trivial because  the light of several components is mixed and
there is a confusion between components in the catalogs.  For 32 stars
identified as evolved, we roughly estimated their masses ${\cal M}_1$
from the absolute magnitudes $V_{abs}$ as ${\cal M}_1 = 0.2(9 -
V_{abs})$.  Those stars in Fig.~\ref{fig:Vabs} that are above the MS
but not marked with asterisks belong to multiple systems that contain
more massive evolved visual components, while the spectroscopic
primaries are unevolved. The median mass of a primary in our sample is
${\cal M}_1 = 1.1{\cal M}_\odot$, the full range is 0.42 to $1.7{\cal
M}_\odot$.

The masses of spectroscopic secondaries ${\cal M}_2$ are calculated
from the known mass ratios in case of double-lined systems (SB2s) or
from the mass functions in case of SB1s (minimum masses).  It is known
that minimum masses are statistically biased (Goldberg et
al. \cite{Goldberg}).  However, we are not concerned with the mass
ratio of SBs, so this bias does not affect the present investigation..

\section{Adaptive-optics observations}
\label{sec:AO}

Adaptive optics (AO) imaging in the near infrared (IR) is a powerful
tool for discovering low-mass companions, as demonstrated by a number
of recent studies (e.g. Shatsky \& Tokovinin \cite{ST02}).  AO allows
for an exploration of a large part of the parameter space that is not
accessible to traditional observational techniques. This increased
capacity results from the high angular resolution, high dynamic range
(the ability to detect faint stars because AO concentrates light into
diffraction-limited cores) and observations in the infrared which
improves the contrast of low-mass companions with respect to their
primary stars.

\begin{figure}[ht]
\centerline{\psfig{figure=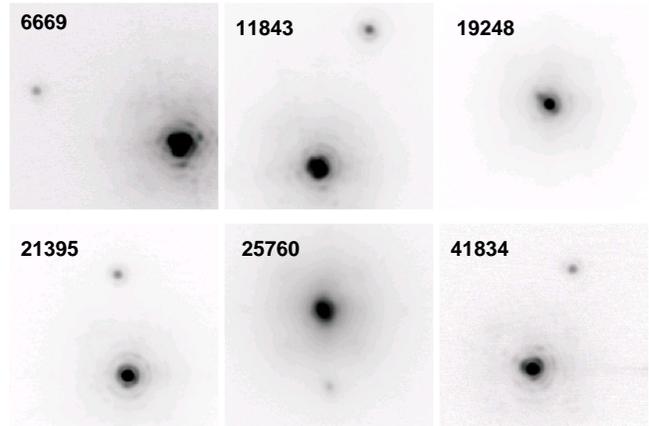,width=8.5cm} }
\caption{Representative narrow-band NACO images of some resolved
targets in $2'' \times 2''$ fields (inverted intensity scale with
square-root stretch). The Hiparcos numbers are marked.
\label{fig:naos} }
\end{figure}

High spatial resolution images of the target stars have been obtained
at the VLT telescope using the NAOS-Conica adaptive optics system,
NACO\footnote{http://www.eso.org/instruments/naco}, on November 8-9,
2004 and on July 6-12, 2005.  We observed 72 targets and two
astrometric calibrators (see below) in the narrow-band (NB)
filter centered at the wavelength 2.12~$\mu$m, to avoid detector
saturation.   Five stars were also observed in the $J$ and/or $H$
photometric bands to estimate the colors of their companions.  Only
targets without known TCs (or with distant tertiaries) were observed.
Of the 72 stars observed, only 62 belong to our final sample. The
remaining ten SBs (HIP 5436, 25760, 34003, 35600, 41834, 50966, 87330,
87370, 110514, 112009) do not satisfy the criteria for inclusion
in the study sample, but we nevertheless provide data on their
companions in the electronic table.

The images were processed in a standard way.  The sky background and
detector bias were estimated and subtracted by median filtering 
a series of five frames dithered on the sky.  The de-biased images
were then flat-fielded and combined.  The relative positions and
magnitudes of wide components were determined by DAOPHOT point spread
function (PSF) fitting of the main star to all sources.  Measurements
on individual images indicate that the random error in position is
about 0.5~mas in each coordinates and the rms error in magnitude
difference is around $0.02^m$ for components with $\Delta m <3^m$,
increasing to 5~mas and $0.05^m$ for $\Delta m = 5^m$.

  Data on close ($<1''$) companions were processed by an
alternative technique of simultaneous fitting of binary parameters and
a non-negative PSF.  First, we determine  the initial binary
parameters in Fourier space by fitting (at high spatial
frequencies above the seeing cutoff) the image Fourier transform (FT)
to the FT of a PSF star convolved with the binary.  This initial
estimate of the binary parameters then serves to de-convolve the
object FT from  the binary, extracting an estimate of the true
(synthetic) PSF.  The synthetic PSF is radially-averaged outside 
the ten pixel (0\farcs13) radius and is used in the second, final
model-fitting.  In this way, it is possible to measure binary
parameters independently of the PSF star.  The quality of this
procedure is controlled by the abscence of  companion traces in
the synthetic PSF before its radial averaging.

Most of  the new companions are well above the detection
threshold (Fig.~\ref{fig:naos}).  In trying to recover faint and
close tertiaries that escape immediate detection, we examined all
images after subtracting radially-averaged profiles.  This procedure
reveals asymmetric features of the PSF, often persistent in the
individual (non-averaged) images. We measured the parameters of eight such
tentative companions and found that their separations are all close to
0\farcs1, with a magnitude difference of about $2^m - 3^m$. The position
angles of these companions differ by $170^\circ$ to $180^\circ$ from
the VLT parallactic angle, strongly suggesting that they are
telescope-related artefacts.  In the case of HIP~107095, a very good
radial-velocity coverage (Fekel \cite{Fekel97}) virtually excludes the
existence of our tentative 0\farcs1 TC, unless its orbital inclination
is very low.  In contrast, the new close companion to HIP~19248 is
real: it does not follow the parallactic-angle dependence and is
confirmed by other techniques.

Relative  coordinates  of  the  components  in  detector  pixels  were
transformed to  on-sky positions using observations of  two known wide
binaries, HIP~108797 and HIP~116737.  The accuracy of this calibration
is entirely  determined by the  known angular separation  and position
angle of these ``calibrators''.   We adopted the resulting pixel scale
of $13.30 \pm 0.01$ mas/pixel  and offset of $-0.4\degr \pm 0.1 \degr$
to the observed angles.   The calibration remained stable between 2004
and 2005 runs.  The data  on newly discovered tertiaries are presented
in Table~\ref{tab:NACO}.   We also include the data  on our calibrator
binaries in  the last rows.  We thus  detected 13 new TCs  (as well as
the known component HIP~98578B) among  62 targets from the main sample
(detection rate  21\%).  For some  new TCs, we were  able to check
the  correspondence of  their colors  with those  expected for  the MS
dwarfs  (photometric confirmation),  while  two TCs  are confirmed  by
proper motions.   We estimate that  all new companions except  two are
physical,  and support  this with  statistical agruments  presented in
Sect.~4.  Comments on selected systems are provided below.

\begin{table}
\caption{Data on  companions  discovered with  NACO.  The columns
give (1) Hipparcos number,  (2) companion identification, (3) spectral
band, (4) epoch of  observation, (5,6,7) measured separation, position
angle and magnitude difference. }
\label{tab:NACO}
\begin{tabular}{c c c c ccc }
\hline
\hline
HIP & Cmp & Band & Epoch & $\rho,$ & $\theta,$ & $\Delta m$ \\
    &       &      & 2000+ & $''$    & $^\circ$  &            \\
\hline
6669 & B & NB   & 4.857 & 1.469 & 69.37 & 5.50 \\
     &   & J    & 4.857 & 1.467 & 70.31 & 6.29 \\
     &   & H    & 4.857 & 1.466 & 70.04 & 5.84 \\
11843 & B & NB  & 4.857 & 1.422 & 339.2 & 3.55 \\
19248 & B & NB & 4.857 & 0.104 & 46.11 & 2.61 \\
      &   & H  & 4.857 & 0.094 & 47.18 & 2.39 \\
21395 & B & NB & 4.857 & 0.991 & 5.15  & 3.61 \\
25760 & B & NB & 4.857 & 0.751 & 184.20 & 4.21 \\
41834 & B & NB & 4.857 & 1.047 & 337.93 & 3.85 \\
      &   & J  & 4.857 & 1.056 & 336.47 & 4.51 \\
      & C & NB & 4.857 & 3.573 & 202.83 & 5.30 \\
43557 & B & NB & 4.857 & 3.446 & 333.73 & 4.28 \\
44164 & B & NB & 4.857 & 1.851 & 270.83 & 6.93 \\
      &   & J  & 4.857 & 1.855 & 270.80 & 7.29 \\
48215 & B & NB & 4.857 & 0.759 & 124.09 & 2.49 \\
64219 & B & NB & 5.533 & 0.309 & 291.41 & 4.05 \\
91360 & Opt& NB& 5.526 & 3.760 & 31.76  & 7.59 \\
94863 & B & NB & 4.857 & 9.378 & 52.20  & 4.68 \\
98578 & C & NB & 5.517 & 0.391 & 340.23 & 2.25 \\
98578 & B & NB & 5.517 & 3.696 & 353.64 & 0.62 \\  
107779& B & NB & 4.857 & 2.219 & 144.60 & 3.56 \\
      &   & J  & 4.857 & 2.226 & 144.59 & 4.16 \\
      &   & NB & 5.522 & 2.233 & 144.30 & 3.45 \\
\multicolumn{6}{c}{Calibrators} \\
108797 & B & NB & 4.857 & 3.848 & 246.37 & 2.52 \\
116737 & B & NB & 4.857 & 3.890 & 276.96 & 0.98 \\
\hline
\end{tabular}
\end{table}

{\em HIP 19248} is an astrometric binary in the Hipparcos catalog.  We
estimate the TC period as $\sim$6~yr,  its mass as 0.3~${\cal
M}_\odot$.  The TC was previously inferred from CORALIE's precision RVs.

{\em  HIP  35487} has  a  known  astrometric  TC with  computed  orbit
($P_3=92$~yr,   estimated  separation   $0\farcs4$,   mass  0.34~${\cal
M}_\odot$) which was not detected with NACO; it  must be a white
dwarf.

{\em HIP  41834} has two companions  in the NACO  images. However, the
fainter companion  C was not detected in the $J$-band, thus  it must be
an optical source.

{\em HIP 91360} has such a faint companion that, if physical, it would
have sub-stellar mass. We consider this compaion an optical source
because the density of background objects near this target is very
high, $N_* = 335$ (Sect.~\ref{sec:2MASS}).

{\em HIP 94863:} a wide companion was seen by luck in the corner of
 a dithered field.  It has been independently identified as 
a 2MASS photometric candidate (Sect.~\ref{sec:2MASS}), with a
consistent position.  The proper motion would have moved it by
$0\farcs7$ in 5~yrs if it were only an optical source, hence we
conclude that it is physical.

{\em HIP  98578} is a visual  triple, with the new  AC pair discovered
here.  We estimate  the period of AC  to be $\sim$100~yr;
it could  be detectable as  astrometric perturbation in the  motion of
AB.  The 2356-yr visual orbit computed for AB (Hopmann \cite{Hopmann})
is premature. Without AO observations, we would wrongly consider the B
component as tertiary.

{\em HIP  107779.} The new companion  is confirmed as  physical by our
second-epoch observation and its $J-K$ color.

\section{Search for companions in 2MASS}
\label{sec:2MASS}

\begin{figure}[ht]
\centerline{\psfig{figure=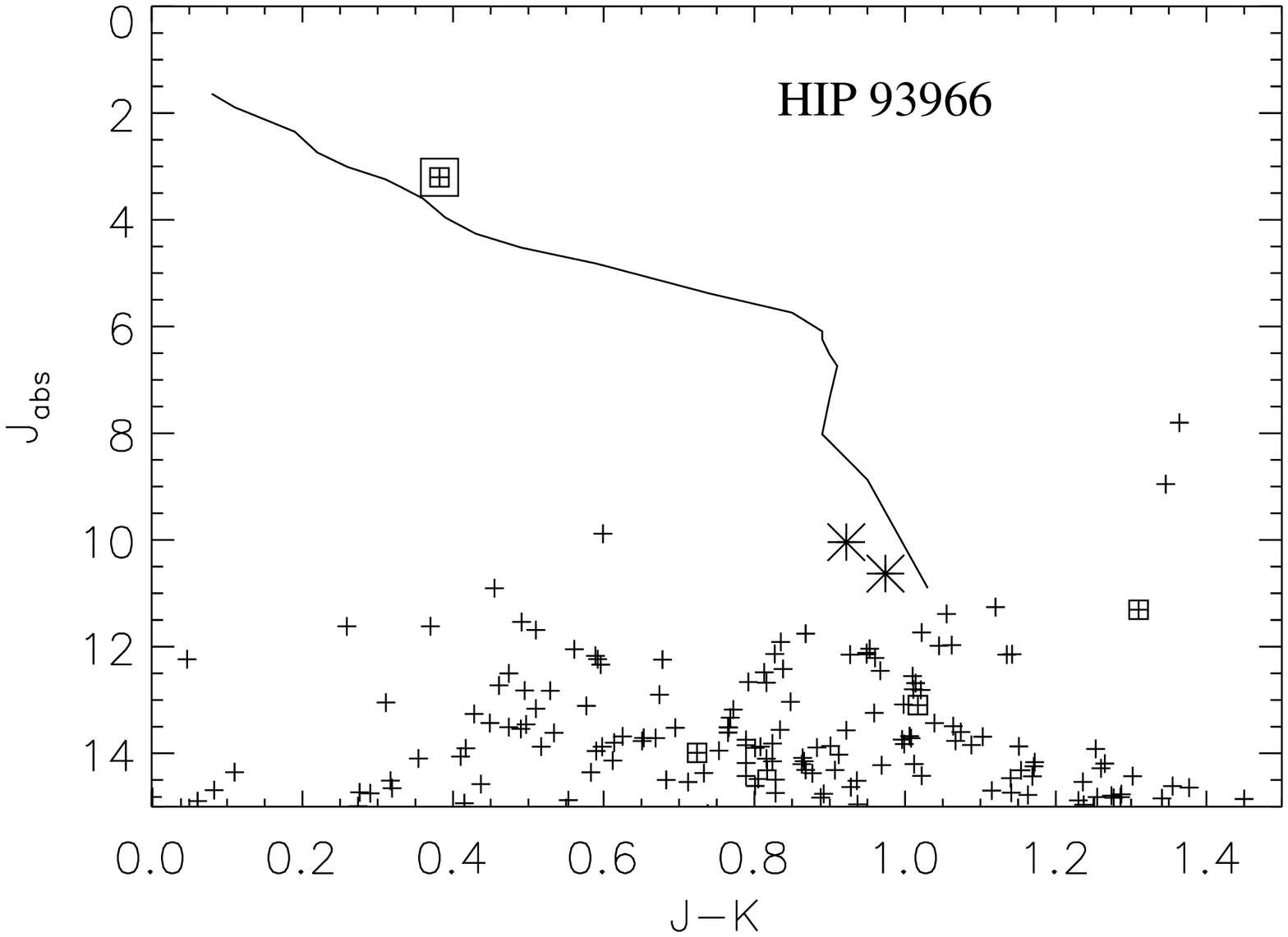,width=8.5cm} }
\centerline{\psfig{figure=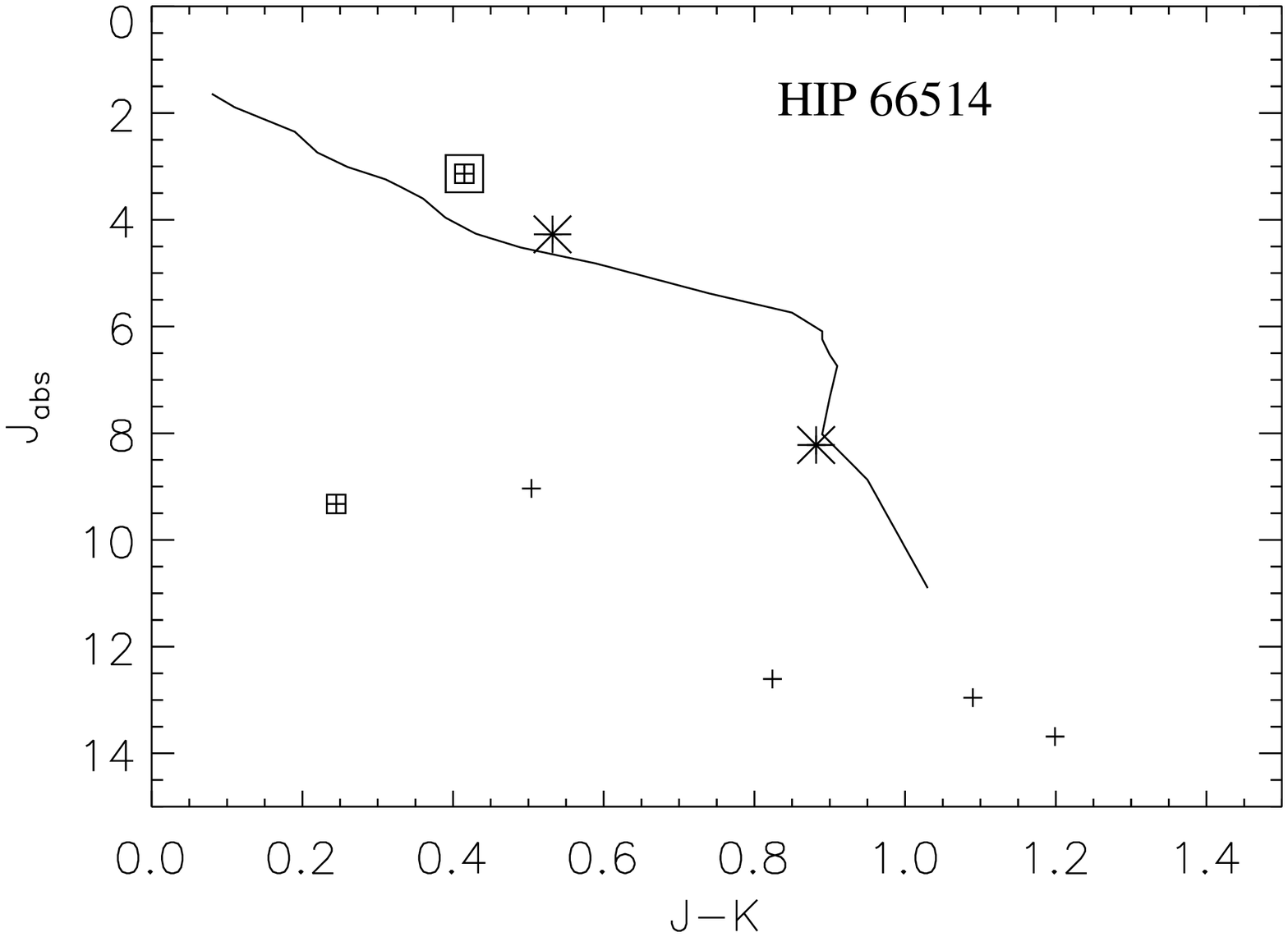,width=8.5cm} }
\caption{CMDs for HIP~93966 in a crowded ($N_*=187$) field (top, two
optical candidates) and HIP~66514 with $N_*=8$ and two physical
companions (bottom).  The primary targets are marked by large squares.
The selected photometric candidates are plotted as large  asterisks,
remaining field stars as crosses.   Stars within $30''$ from the
target are marked by small squares. The  solid line depicts 
the standard Main Sequence from Lang (\cite{Lang}).
\label{fig:CMD} }
\end{figure}

\begin{figure}[ht]
\centerline{\psfig{figure=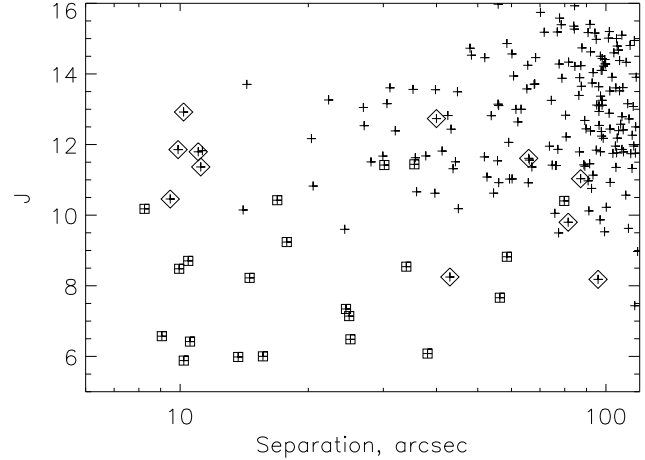,width=8.5cm} }
\caption{Distribution of photometric candidates from 2MASS in 
separation $\rho$ and  magnitude $J$.  Known physical companions
are marked by squares, the new physical companions by diamonds, while
crosses are likely optical (i.e. not physical) companions.
\label{fig:2Mstat} }
\end{figure}

The SBs were examined for the presence of wide visual companions using
 the Two Micron All-Sky Survey (2MASS) and the Digital Sky Survey
(DSS).  The primary targets are bright nearby dwarfs.  Many faint
``companions'' are found near any bright star, but the vast majority
of such companions are simple line-of-sight projections or ``optical''
sources. However, faint {\it physical} companions may still hide among
the multitude of optical companions.  Previous generations of
double-star observers  were guided by statistics to separate
likely physical companions from chance projections (Poveda et
al. \cite{Poveda}).  Thus, a ``filter'' was adopted to reject optical
companions, and such  a filter introduces a strong selection
effect in the existing double-star catalog, WDS (Mason et al.
\cite{WDS}).  In this study, we use the $JHK_s$ photometry provided by
2MASS to place the companion stars on the color-magnitude diagram
(CMD) and to select potential physical companions.   The
extinction in the $J$ and $K$ bands can be safely neglected for our
nearby sample.

The photometry and coordinates of all stars within a $2'$ radius from
each of the 165 targets were retrieved from the 2MASS Point Source
Catalog using the Vizier service at CDS.  A total of 6079 companions
were thus found.  Only 202 (3.3\%) were selected automatically as {\it
photometric candidates} by their distance from the MS $\sqrt{ \Delta
J^2 + \Delta(J-K)^2} <0.2^m$.  We checked that this criterion selects
most of the known physical companions, but it cannot be made
``sharper'' without  the risk of losing companions.  Two examples
of CMDs are shown in Fig.~\ref{fig:CMD}.  It is clear that in the
fields with  a large number of stars $N_*$ the confusion impedes
 the detection of physical companions and that the photometric
candidates in these fields are likely optical.

The statistics of photometric candidates (Fig.~\ref{fig:2Mstat})
confirms that they are mostly optical -- faint and with separation
$\rho$ of the order of  the field  radius, $120''$. In
contrast, known physical companions are brighter and are distributed
in $\log \rho$ almost uniformly.

In an effort to reduce the number of candidates, we discarded 36
fields with $N_*>40$ (the median is $N_*=13$).  Faint companions with
$\rho>60''$ and $J>13^m$ were also discarded, as well as multiple
candidates and candidates in the Hyades (HIP 20019, 20284, 20440).
Excluding known physical companions, we retained 35 candidates for
astrometric checking.

The targets are nearby stars with typical proper motions (PMs) of
$0\farcs1$ per year.  Physical companions must share these motions and
would be displaced by $\sim 5''$ on the sky in 50~yr.  Old
photographic plates are over-exposed in the vicinity of  the
targets, hence we can check only candidates with $\rho >20''$.  The
images of fields around each target were retrieved from the DSS (red
Palomar plates for northern declinations and red UK Schmidt plates for
southern declinations). The time base between 2MASS and DSS is around
50~yrs in the North but only around 10~yrs for southern stars.

We found  photometric candidates  in the DSS  images and   roughly
measured their  coordinates using  the astrometric reference  of the DSS,
with an  accuracy of about  $0\farcs5$.  When the displacement  of the
main  target  is  significantly   larger  than  $0\farcs5$,  we  can
discriminate the  optical companions.  The majority of  wide and faint
companions turned  out to be  optical.  In four  cases the status  of the
companions remained uncertain (small PM and/or small time base), while
three new companions with large separations (HIP 16042, 36328, 66514) 
among our 35 selected candidates were confirmed.

The distribution of new companions in the $(\log \rho, J)$ plane
(Fig.~\ref{fig:2Mstat}) shows a ``cluster'' of 5 points near $\rho
\approx 10''$.  These TCs have not been checked by astrometry.
However, statistical arguments show that they are all likely physical
companions.  We  leave out HIP~94863 which lies in a crowded field
($N_* = 52$), but note that its $9\farcs4$ companion is physical
(Sect.~\ref{sec:AO}).  The remaining 4 stars with such companions are
HIP 3362, 17076, 20712, 60331. These targets have small stellar
surface densities ($N_*$ of 17, 13, 15, 6, average 12.75) and the
companion separations are less than $11''$.  Among 86 targets with
$N_*< 13$, the expected number of point sources at $\rho <11''$ is
8.7.  Given that the photometric criterion selects only $\approx$3\%
of all sources, the expected number of randomly selected photometric
candidates within $11''$ is 0.26; this number should be compared to
the four companions actually found. In addition, there are 6
previously known {\em physical} TCs with similar separations  of
$\rho \sim 10''$.

An additional screening for new close companions has been performed by
examining the 2MASS images of apparently single targets. We found such
companions to HIP~43557 (confirmed with NACO) and HIP~60956. The
latter is still considered as tentative.

 Summarizing the results for 2MASS candidates,
Table~\ref{tab:2MASS} lists the data on 23 known, 7 new certain, and 5
new tentative (marked with ?)  companions  -- a total of 35 TCs.
The known companions are marked ``MSC''.  We see that existing
catalogs contain about 2/3 of the wide  TCs; another
1/3 are added in this work.

\begin{table}
\caption{Tertiary companions found in 2MASS.  The columns give (1) HIP
number, (2,3) separation and position  angle, (4) $J$ magnitude of the
companion, (5) $J-K$ color, (6)  status, with MSC for previously known
companions. }
\label{tab:2MASS}
\begin{tabular}{c  cccc l }
\hline
\hline
    HIP & $\rho,''$ & $\theta, ^\circ$ &  $J$ &  $J-K$ & Status \\
\hline
   3362 &  11.18  &  278.13  &  11.366  &   0.801  &  new  \\
   7874 &  10.45  &   31.34  &   8.705  &   0.749  &  MSC  \\
  12189 &  38.11  &  274.44  &   6.080  &   0.258  &  MSC  \\
  16042 &  95.82  &  129.08  &   8.182  &   0.556  &  new,CPM  \\
  17076 &   9.89  &  245.34  &  11.857  &   0.839  &  new  \\
  20712 &  11.03  &  119.67  &  11.795  &   0.861  &  new  \\
  24663 &   9.95  &  148.60  &   8.481  &   0.460  &  MSC  \\
  31850 &  30.17  &  248.74  &  11.420  &   0.840  &  MSC  \\
  36238 &  43.00  &  323.59  &   8.249  &   0.780  &  new,CPM  \\
  41211 &  35.49  &  328.69  &  11.441  &   0.841  &  new?  \\
  42172 &  10.20  &   25.11  &   5.882  &   0.385  &  MSC  \\
  45957 &   8.25  &  118.02  &  10.179  &   0.787  &  MSC  \\
  47053 &  24.95  &  149.20  &   7.142  &   0.271  &  MSC  \\
  52064 &   3.67  &  342.14  &   6.136  &   0.034  &  MSC  \\
  56809 &   9.06  &  248.40  &   6.573  &   0.615  &  MSC  \\
  56960 &  79.81  &  334.28  &  10.403  &   0.705  &  MSC  \\
  60331 &  10.19  &  320.43  &  12.921  &   0.896  &  new  \\
  60956 &  5:     &  355:    &   -      &   -      &  new? \\
  61910 &  58.52  &  228.80  &   8.822  &   0.444  &  MSC   \\
  66514 &  56.34  &  258.63  &   7.662  &   0.532  &  MSC  \\
  66514 &  65.86  &  254.25  &  11.610  &   0.882  &  new,CPM  \\
  73269 &  81.52  &   39.64  &   9.800  &   0.716  &  new?  \\
  74037 &  17.80  &   19.75  &   9.239  &   0.494  &  MSC  \\
  84586 &  33.97  &   92.78  &   8.542  &   0.913  &  MSC  \\
  91009 &  16.91  &   45.50  &  10.424  &   0.902  &  MSC  \\
  94863 &   9.48  &   53.03  &  10.457  &   0.769  &  NACO  \\
  98578 &   3.00  &  353.70  &   6.251  &   0.103  &  MSC  \\
 103569 &  10.55  &   67.95  &   6.419  &   0.283  &  MSC  \\
 104026 &  39.99  &  201.07  &  12.735  &   0.865  &  new?  \\
 107354 &  14.56  &  289.05  &   8.224  &   0.572  &  MSC  \\
 108461 &  13.69  &  246.71  &   5.984  &   0.379  &  MSC  \\
 111802 &  24.52  &  350.36  &   7.344  &   0.853  &  MSC  \\
 114379 &  15.65  &   75.94  &   6.002  &   0.332  &  MSC  \\
 114639 &  87.20  &  205.54  &  11.033  &   0.805  &  new?  \\
 114703 &  25.12  &  355.67  &   6.484  &   0.338  &  MSC  \\
\hline
\end{tabular}
\end{table}

\section{Parameters of tertiary companions}
\label{sec:par}

\begin{figure}[ht]
\centerline{\psfig{figure=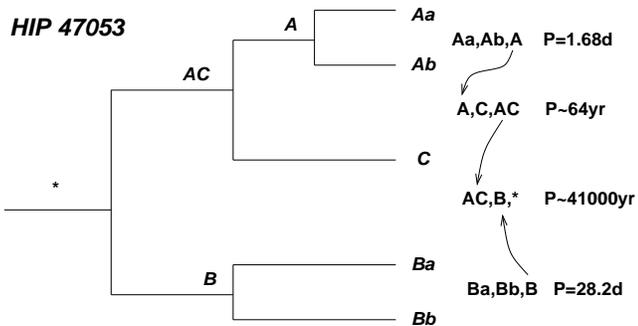,width=8.5cm}}
\caption{Example of the hierarchical  quintuple system HIP~47053 and the
 identifications of its sub-systems  (see text).
\label{fig:tree} }
\end{figure}

\begin{table*}[ht]
\caption{Data on  companions (fragment). The columns  are explained in
  the text.}
\label{tab:comp}
\begin{tabular}{c l l r l r l r r l }
\hline
\hline
    HIP & ID  &  Type &  \multicolumn{2}{c}{Separation}   &  \multicolumn{2}{c}{Period}  & ${\cal M}_1$  & ${\cal M}_2$ &    Remark \\
\hline
 47053 & Aa,Ab,A &    S2 &  0.471 &mas &   1.681 &d &1.54 a &1.46 a &             SB 575 \\
 47053 &  A,C,AC &     v &  0.300 &"   &  64.257 &y &3.00 s &0.98 a &           COU 2084 \\
 47053 &  AC,B,* &  Chrp & 25.000 &"   &  41.099 &ky&3.98 s &1.65 s &           STF 1369 \\
 47053 & Ba,Bb,B &    S1 &  2.530 &mas &  28.231 &d &1.29 a &0.36 m &1998AstL...24..288T \\
 48215 & Aa,Ab,A &    S1 &  0.924 &mas &   3.390 &d &1.11 a &0.29 m &      SB 584 DI Leo \\
 48215 &   A,B,* &    AO &  0.759 &"   & 130.282 &y &1.40 s &0.61 v &          NACO-2004 \\
 48273 & Aa,Ab,* &    S2 &  1.196 &mas &   3.055 &d &1.21 a &1.15 q &             SB 585 \\
 48833 & Aa,Ab,* &    S2 &  1.057 &mas &   3.055 &d &1.26 a &1.20 q &             SB 584 \\
 49018 & Aa,Ab,A &  E,S2 &  0.690 &mas &   1.070 &d &0.78 a &0.53 q &     DH Leo, SB 591 \\
 49018 &   A,B,* &     v &  0.220 &"   &  14.190 &y &1.31 s &0.50 v &          CHARA 145 \\
 49161 & Aa,Ab,* &    S1 &  3.965 &mas &  30.000 &d &0.74 a &0.50 q &         CORALIE     \\
 49809 & Aa,Ab,* &    S1 &  5.557 &mas &  28.098 &d &1.49 a &0.20 m &             SB 601 \\
\hline
\end{tabular}
\end{table*}

 Table~\ref{tab:comp} contains the data on previously known
companions to the SBs which have been extracted from the MSC
(Tokovinin \cite{MSC}) and combined with the new detections in a
single database.  Additional bibliographic searches for stars without
reported tertiaries have been made through SIMBAD. We also scanned the
fourth catalog of interferometric measurements (Hartkopf et al.
\cite{INT4} -- INT4) and marked the stars that were observed by
speckle interferometry.  Also, the ninth catalog of spectroscopic
binary orbits (Pourbaix et al. \cite{SB9} -- SB9) was searched for new
orbits.

A  part of  Table~\ref{tab:comp}  with the  information  on all  known
components   is   printed   here;   the  full   Table   is   available
electronically. Below  we describe  the information   in this
Table.  Additional comments on  specific systems are provided in the
electronic notes to Table~\ref{tab:comp}.

Identification  of each  system is  given in the  column 2  as a
sequence of  3 designations (primary, secondary,  parent) separated by
commas.   These  codes  describe  the  hierarchy  of  each  system  by
referencing    to        the    parent,   as    illustrated    in
Fig.~\ref{fig:tree}.  The system at the highest hierarchy level (root)
is coded with an asterisk,   it is the parent of the wide sub-system
AC,B in Fig.~\ref{fig:tree}.

The type code  of each system shows the  method(s) of its discovery,
in the same manner as in the  MSC.  For example, {\bf  S1} and {\bf S2}
refer  to  single-  and  double-lined  SBs, {\bf  E}  to  eclipsing
systems, {\bf  C} stands for  wide (CPM) systems (the  following small
letters   describe  which  criteria   of  physical   relation  between
components are  satisfied,   see MSC and  notes to  the electronic
Table~5), {\bf v} means a resolved system closer than
$3''$, {\bf  V} stands for a  visual binary with  computed orbit, etc.
Two new special  types are {\bf AO} (companions  discovered with NACO)
and {\bf 2M} (2MASS companions).

The separation between components is  given together  with 
units (arcseconds or milliarcseconds).   For systems with known orbits
the  separation refers to semi-major  axis, otherwise  it is  the observed
separation  $\rho$. The  separation of SBs is  estimated  from known
periods and Kepler's Third Law as

\begin{eqnarray}
\rho \approx p P^{2/3} {\cal M}^{1/3},
\label{eq:sep}
\end{eqnarray}
where $p$ is  the parallax (arcseconds), $P$ is  the orbital period in
years and the  system mass ${\cal M}$ is in solar  units.  Here we assume
implicitly that  separation is statistically  equivalent to semi-major
axis.

Periods are  given in the next column, in  units of days, years
or kiloyears. For wide systems without computed orbits the periods are
estimated from $\rho$ with  Eq.~\ref{eq:sep}.

Masses of  primary and  secondary components  ${\cal  M}_1$ and
${\cal M}_2$ are  given in solar mass units  with codes indicating the
method of mass estimation.  The  preferred methods are {\bf a} -- from
spectral type (cf.  Sect.~\ref{sec:sample}) or {\bf *} -- estimated by
other authors  from orbit solutions  or detailed models.   Other codes
are {\bf l}  -- evolved components, {\bf s}  -- sum of sub-components'
masses,  {\bf q}  -- SB2  secondary, {\bf  m} --  minimum mass  of SB1
secondary,  {\bf v}  -- mass  from  magnitude difference,  {\bf :}  --
unknown  magnitude  difference,  {\bf  ?}   --  uncertain  companions.
Masses of all new components (AO and 2M types) were estimated from the
magnitude differences in the $K$  band using the relations of Henry \&
McCarthy  (\cite{HM93})  and taking  into  account  the  light of  the
spectroscopic secondary.

Remarks  provide  common  identifications for  the  sub-systems,
e.g. the SB  numbers in Batten et al.   (\cite{SB8}) catalog, visual
  double-star designations, bibcodes, etc.  

The sample contains  165 spectroscopic binaries belonging to 161
stellar systems; 79 have no known tertiaries, and the remaining 86 SBs
have one or  more companions (64 triples, 11  quadruples, seven quintuples
 -- a  total of 82 systems).  Of these,  six TCs are
uncertain (our  2MASS detections, doubtful  speckle companions, etc.),
but they are still considered  in the statistics.  The fraction of SBs
with  TCs  should  be no  less  than  86/165=0.52.   Not all  TCs  have been
discovered   yet.   Known   distant  companions   are   considered  as
tertiaries,  but  in  some  systems  the  true  tertiaries  at  closer
separations  are not  known  and the  present ``tertiaries''  are
actually higher-level  companions, as  turned out to  be the  case for
HIP~98578, where  a new,  closer companion to  SB was  discovered
with NACO in addition to the known wide companion.

Mass ratios of TCs are defined here as $q_3 = {\cal M}_3/{\cal
M}_1$, i.e.  relative to the spectroscopic primary, not to the total
mass of the SB.  There is no ambiguity for triples, but for systems
with more components, as in Fig.~\ref{fig:tree}, the choice of
tertiary is not always evident.  We select ${\cal M}_3$ to be the
largest mass of all components at the next higher level of hierarchy.
Thus, the binary Ba,Bb in  Fig.~\ref{fig:tree} has Aa as its most
massive tertiary, $q_3 = 1.10/1.29 = 0.85, \;\; P_3 \approx 41000$~yr.
However, another close binary Aa,Ab in the same system has  a
tertiary component C, hence $q_3 = 0.98/1.10 = 0.89, \;\; P_3 \approx
64$~y.

\section{Detection limits}
\label{sec:det}

\begin{figure}[ht]
\centerline{\psfig{figure=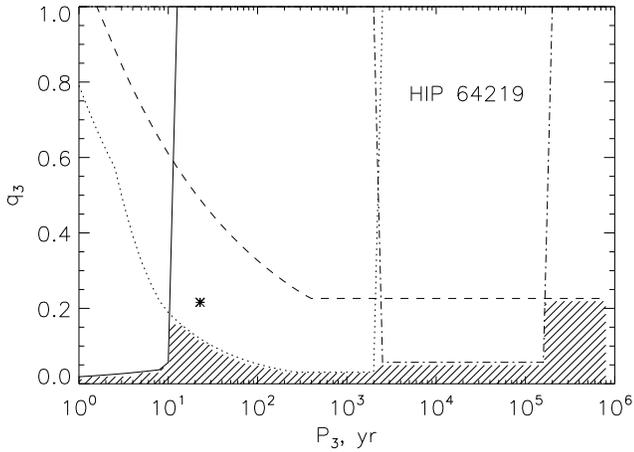,width=8.5cm}}
\caption{Estimated   limits  of   companion  detection   by  different
  techniques for  HIP~64219. Full  line -- radial  velocities, dotted
  line --  NACO, dashed  line -- visual  and CPM, dash-dotted  line --
  2MASS. The NACO companion is marked by asterisk.   The overall
  probability of TC detection for this object is zero in the hatched zone below
  all curves and one above. 
\label{fig:exam} }
\end{figure}

\begin{figure}[ht]
\centerline{\psfig{figure=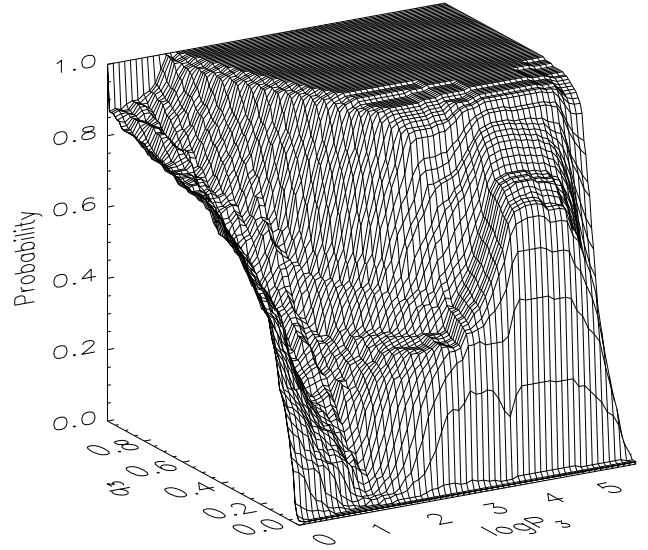,width=8.5cm}}
\caption{Probability of tertiary companion detection for the whole
  sample as a function of $\log P_3$ and $q_3$.
\label{fig:qdens} }
\end{figure}

For any  given star  and any observing  technique, the  probability of
detecting  a TC  depends on   the  companion's parameters  -- its
period, $P_3$,  and its mass ratio, $q_3$. This probability  should be a
smooth  function of the parameters,  but we  simplify it  here as  a sharp
limit,  assuming  that all  TCs  with  $q_3  > q_{\rm  lim}(P_3)$  are
detected and less massive companions are not.  An example of such
limits  for  one target  is  shown  in  Fig.~\ref{fig:exam}.  By  
averaging the detection probabilities for the whole sample, we 
again  obtain a  smooth  detection probability  (Fig.~\ref{fig:qdens}).   The
method  of computing detection  limits is  detailed in  Appendix~A for
each observing  technique.  We specify  the detection limits  for each
{\em system},  hence the  four systems  with two SBs  each have  only one
entry per system.   However, these systems are at  least quadruple and
pose no problem in the statistics.

\begin{figure}[ht]
\centerline{\psfig{figure=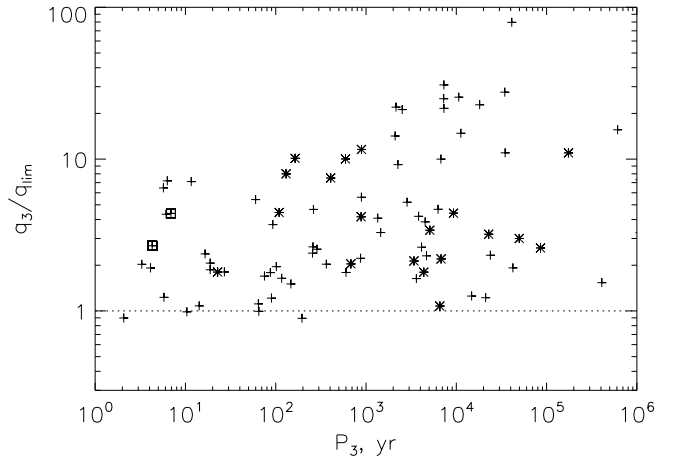,width=8.5cm}}
\caption{ Comparison of the  actual mass ratios of tertiary components
$q_3$ with  their estimated detection limit $q_{lim}$.   Known TCs are
plotted   as    pluses,   the   new   (NACO   and    2MASS)   TCs   as
asterisks. Uncertain TCs are marked with squares.
\label{fig:chklim} }
\end{figure}

Detection  limits  were  converted   into  estimates  of  the  minimum
observable mass ratio $q_3$  (Sect.~\ref{sec:par}) and compared to the
actual    mass   ratios    of    known   and    new   TCs    in
Fig.~\ref{fig:chklim}.  We  see   that  only  several  tertiaries  are
slightly  below the estimated  detection limit  and conclude  that the
limits can be used confidently in the statistical analysis.

\section{Statistics of tertiary companions}
\label{sec:stat}

\subsection{Mass ratios and periods of tertiaries}

\begin{figure}[ht]
\centerline{\psfig{figure=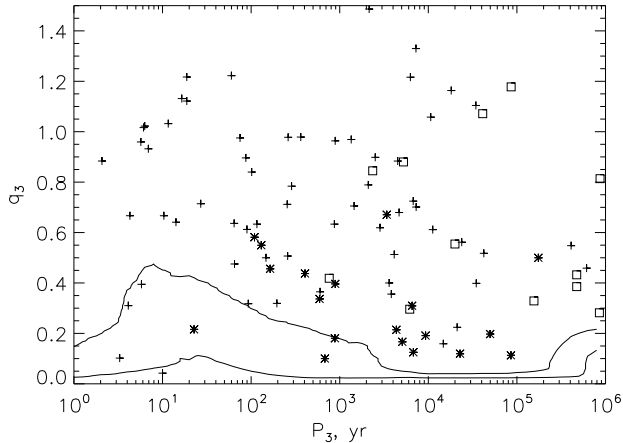,width=8.5cm} }
\caption{Mass  ratio of tertiary  to spectroscopic  primary   as a
function of tertiary orbital period.  The known components are plotted
as  pluses,  the  new  components  as asterisks.  The  squares  denote
components at the next hierarchical  level, not considered in the main
statistical analysis.   The lines  trace  detection probabilities  of   10\%
(lower)  and  50\%(upper),  i.e.   the  contours  of  the  surface  in
Fig.~\ref{fig:qdens}. 
\label{fig:q3} }
\end{figure}

Figure~\ref{fig:q3}  shows the  distribution of  TCs  in $(P_3,
q_3)$ space. Known companions  at higher hierarchical levels (like the
component Ba relative to  the Aa, Ab system in Fig.~\ref{fig:tree}) are
also plotted as squares  for comparison. The curves indicate detection
limits. As expected, new TCs  detected in this work mostly have 
low mass ratios.

We note  that 16 tertiaries  ($17 \pm 4$\%)  have $q_3 > 1$,  i.e. the
spectroscopic primary is not the most massive component in the system.
If components  were selected randomly, one would  expect this fraction
to be 1/3 or higher (in systems with more than 3 components we select
the most massive tertiary). Hence,  we confirm the known tendency that
the  most massive  components in  multiple systems  are preferentially
found in  close sub-systems.

Interestingly, we  find only three TCs  with $P_3 > 10^5$~yrs,  but also six
higher-level companions  wich such long periods.  Thus,  a decrease of
the  number of  TCs at  long  periods must  be real,  not a  selection
effect.

\subsection{SBs with and without tertiary companions}

\begin{figure}[ht]
\centerline{\psfig{figure=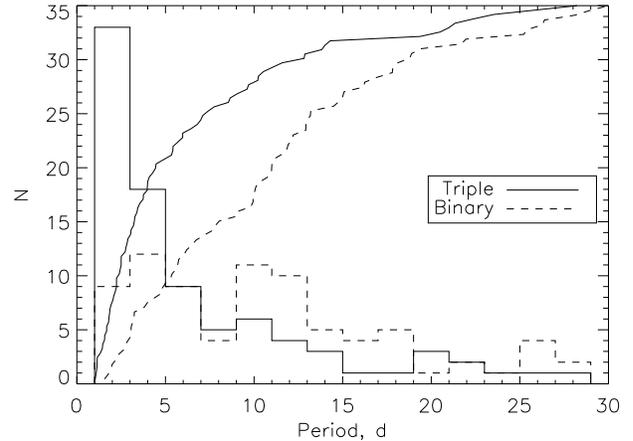,width=8.5cm}   }
\caption{Histograms and cumulative distributions of orbital periods of
spectroscopic binaries with tertiaries (Triple, full line) and without
tertiaries (Binary,  dashed line).   The larger fraction  of short
periods among those SBs which belong to triples is evident.
\label{fig:p-p} }
\end{figure}

The cumulative distributions of the orbital periods of SBs with and
without TCs are compared in Fig.~\ref{fig:p-p}.  It is evident that
SBs within multiple systems have, generally, shorter orbital periods.
The maximum difference between the cumulative distributions 
normalized to 1 reaches 0.30.  The numbers of objects in the
histograms are 86 and 79, so the Kolmogorov-Smirnov test rejects the
hypothesis that both distributions are equal with a significance level
of 0.999.  Some of the SBs considered today as ``single'' have yet
undiscovered tertiaries (e.g.  suspected astrometric companions), so
the actual difference between the distributions is  likely to be
larger.  The presence of a TC shortens the
period of a spectroscopic binary.  On the other hand, the existence of
such difference shows that SBs without TCs (pure binaries) do exist.

\begin{figure}[ht]
\centerline{\psfig{figure=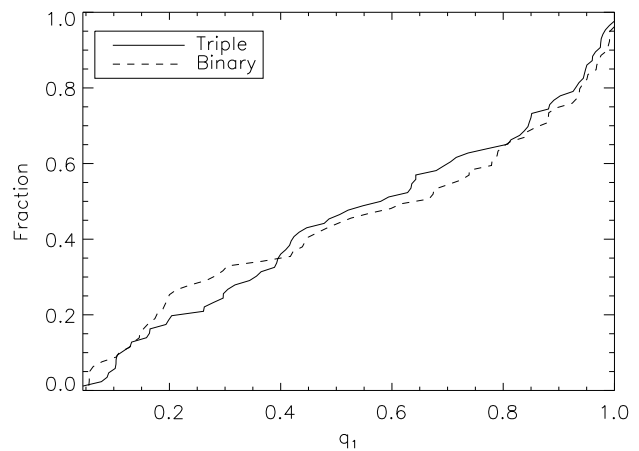,width=8.5cm}   }
\caption{Cumulative distributions of  the mass ratios of spectroscopic
binaries with  tertiaries (Triple,  full line) and  without tertiaries
(Binary, dashed line).
\label{fig:p-q} }
\end{figure}

In contrast, there is no difference in the mass ratio distributions of
the SBs with and without TCs: both show a slight excess of systems
with nearly equal-mass components (twins), but otherwise the
distributions are almost uniform (Fig.~\ref{fig:p-q}).  As noted in
Sect.~\ref{sec:sample}, the mass ratios  of SB1 are biased, but
this bias is independent of the presence of the TC.

\subsection{Period-period diagram}

\begin{figure}[ht]
\centerline{\psfig{figure=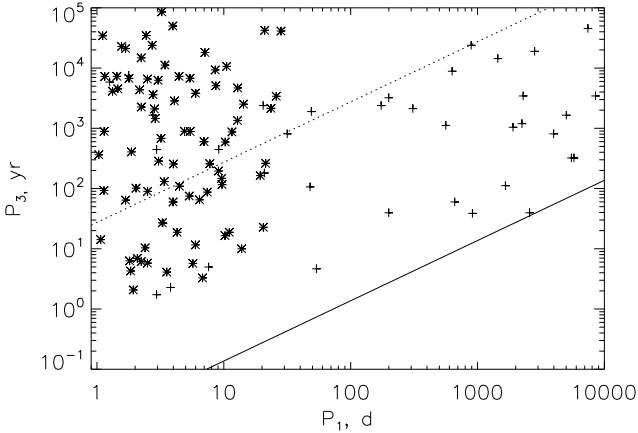,width=8.5cm}   }
\caption{Comparison of the SB periods (horizontal axis) with
  respective tertiary periods (vertical axis). The full line denotes
  the dynamical stability limit $P_3 = 5 P_1$, the dotted line marks
  the period ratio of 10\,000. Our sample is plotted as asterisks,
  other triple systems from MSC are plotted as crosses.
\label{fig:p1p3} }
\end{figure}

There seems to be no correlation between the SB orbital periods $P_1$
and the TC periods $P_3$ (Fig.~\ref{fig:p1p3}, asterisks).  However,
the shortest TC periods (lower boundary of points) seem to increase
with $P_1$.  For comparison, we overplot nearby solar-type multiples
from the MSC (crosses, cf.  Tokovinin \cite{Merida}).  Multiple
systems with wide inner binaries are almost exclusively found with
period ratios $P_3/P_1$ of between 5 and 10\,000.  In contrast, our
multiples are {\em all} quite far from the dynamical stability limit
and can have period ratios $>10^4$.  This difference may be understood
if the SBs were formed with longer periods and later migrated to shorter
periods.  The relative paucity of inner sub-systems with periods of
a few months matches this scenario, although it could probably also be
explained by observational selection.  The lowest observed $P_3 \sim
2$~yr implies that the progenitors of these SBs had $P_1 \le P_3/5
\sim 150^d$, otherwise those systems would have disintegrated 
through dynamical instability (e.g. Mardling \& Aarseth
\cite{Mardling}).

\subsection{Companion distribution by Maximum Likelihood}.

An estimate of  the TC distribution over period  $P_3$ is derived here
by the  Maximum Likelihood  (ML) method. The parameter  space is
divided into bins, the fraction of TCs  in each bin is ${\bf f} = \{
f_k \}$.  The  probability of obtaining the actual  data ${\cal L}$ is
called the likelihood.  By minimizing its natural logarithm $S$

\begin{equation}
S({\bf f}) = -2 \ln {\cal L}({\bf f}),
\label{eq:S}
\end{equation}
 we find the  estimate of ${\bf f}$.  Moreover,  the ML method permits
us to determine   confidence intervals of the  estimated parameters or
their functions  (Avni et al. \cite{Avni}).   Mathematical details and
ML equations are given in Appendix~B.

In the analysis, we select five bins for $P_3$ uniformly covering the
$\log P_3$ space from one to $10^6$~yrs.  The detection probabilities
are computed as a surface above the $q_{\rm lim}(P_3)$ curve in each
period bin. This is strictly true only for a uniform distribution of
$q_3$.  The actual $q_3$ distribution does in fact appear rather uniform
(Fig.~\ref{fig:q3}).  A more correct analysis of the joint TC
distribution in $P_3, q_3$ (with 3 bins in $q_3$) has been performed
and gives practically identical results.

\begin{figure}[ht]
\centerline{\psfig{figure=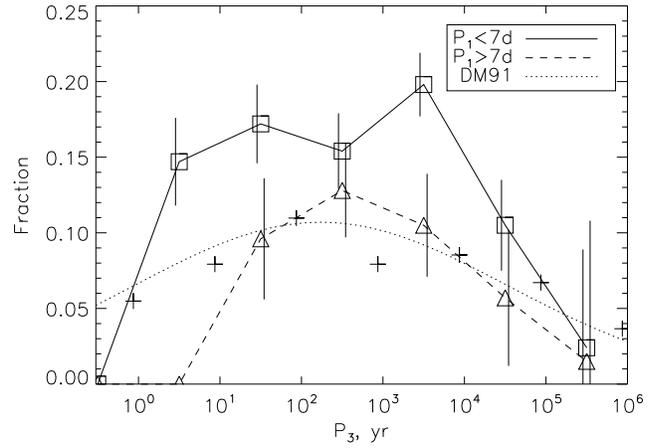,width=8.5cm} }
\caption{Period distributions of  TCs around close SBs
(squares) and wide SBs (triangles).  The error bars correspond to $\pm
1  \sigma$.  The  distribution  of solar-type  binaries  from DM91  is
plotted for comparison (dotted line and crosses).
\label{fig:dm91} }
\end{figure}

 The total TC fraction  estimated by ML, i.e. corrected for
incompleteness, is $63 \pm 5$\%.  Considering the difference between
short- and long-period SBs, we split our sample in two roughly equal
parts: close ($P_1 < 7$~d, 90 systems) and wide ($P_1 > 7$~d, 75
systems), and repeat the ML analysis for these sub-samples.  The
results are shown in Table~\ref{tab:7d} and in
Fig.~\ref{fig:dm91}. The errors of the TC frequency are determined by
the shape of the $S$-function near its minimum, as illustrated in
Fig.~\ref{fig:S}.  We also provide in Table~\ref{tab:7d} the raw
(uncorrected) TC frequencies $f_{\rm raw}$ to show that our
correction for incompleteness is not dramatic.  The calculation
was repeated ignoring uncertain TCs, and only a slight
reduction of the estimated TC frequency results.

\begin{table}[ht]
\caption{Frequency of tertiary companions  in two sub-samples.  $N$ is
the  number  of  systems in  each  sample,  $f_{\rm  raw}$ is  the  TC
frequency  uncorrected  for  incomplete  detections,  $f$  is  the  TC
frequency  estimated by  the ML  method.  The  numbers in  italics are
obtained by discarding uncertain companions. }
\label{tab:7d}
\medskip
\begin{tabular}{ l | ccc  }
\hline
\hline
Sample & $N$ & $f_{\rm raw}$ & $f$ \\
\hline
$P_1 <7$d & 90 & 0.66 & 0.80 $\pm$ 0.06 \\
          & {\it 88} & {\it 0.65} & {\it 0.79}  $\pm$ {\it 0.06} \\
$P_1 >7$d & 75 & 0.33 & 0.40 $\pm$ 0.06 \\
          & {\it 71} & {\it 0.30} & {\it 0.36}   $\pm$ {\it 0.06} \\
All       & 165& 0.52 & 0.63 $\pm$ 0.05 \\
\hline
\end{tabular}
\end{table}

\begin{figure}[ht]
\centerline{\psfig{figure=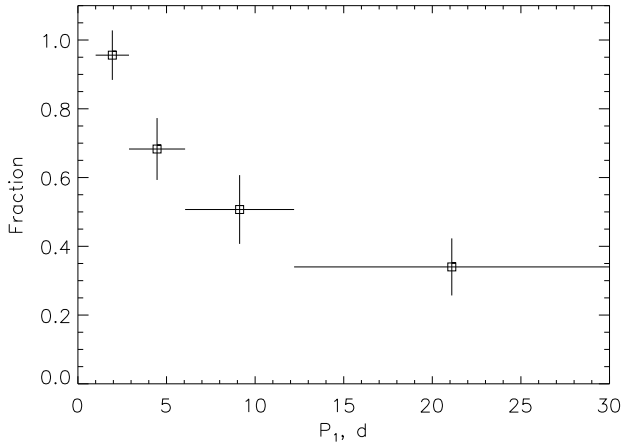,width=8.5cm} }
\caption{Frequency of tertiary companions as a function of SB period.
\label{fig:f3p1} }
\end{figure}

The sample was then further sub-divided into four groups ranked over
SB periods, and the calculation of the TC frequency was done for each
group separately (Fig.~\ref{fig:f3p1}).  For the shortest periods,
$P_1 < 2.9^d$, the estimated TC frequency is as high as 96\%.

It is clear that many SBs have tertiary companions.  But 
companions are typical also for normal solar-type  field stars
(Duquennoy \& Mayor \cite{DM91}).  Is there any difference between SBs
and field stars  with respect to wide companions?  We compare in
Fig.~\ref{fig:dm91} the companion frequencies in these samples, both
corrected for selection effects.  Overall, the companion frequency in
the DM91 sample for periods above one~yr is 50\% -- higher than in the
wide-SB sub-sample.  On the other hand, the fraction of companions to
close SBs is clearly increased. Their periods range from a few
years to several thousand years.

\subsection{High multiplicities}

Our sample contains  64 triples and many systems of multiplicity
higher than 3: 11 quadruples and seven quintuples (Sect.~5).  Batten
(\cite{Batten}) introduced $f_n$ -- the fraction of systems of
multiplicity $n$ and higher to systems of multiplicity $n-1$.  For our
sample, $f_4 = (11+7)/64 = 28$\% and $f_5 = 7/11 = 64$\%.  These
numbers are comparable to the high-multiplicity frequencies in the
MSC: $f_4 = 179/626 = 29$\% and $f_5 = 38/141 = 27$\% (Table~1 of
Tokovinin \cite{T2001}).  Spectroscopic binaries from the MSC, such as
HIP~76563/76566 (two SBs of 3.27$^d$ and 14.28$^d$ in a system with
six components), are not included in the sample. This is done to avoid
intentional bias towards high multiplicity.

\section{Summary}
\label{sec:sum}


\begin{enumerate}
\item
The period distributions of SBs with and without tertiary companions
(TCs) are significantly different. This proves the existence of pure
SBs without TCs.

\item
The mass ratio distributions of SBs with and without TCs are
 identical.

\item
The frequency of TCs is 63\% for the whole  SB sample. However,
it is a strong function of the SB period, reaching 96\% for the close
($P_1<3^d$) SBs and decreasing to 34\% for SBs with $P_1>12^d$
(Fig.~\ref{fig:f3p1}).  The last number is less than the companion
frequency  of $\sim$50\% for G-dwarfs in the field.   These
results are  robust because the TC detection does not depend on the
SB period.

\item
The  periods of   most TCs  in  our sample  range  from 2~yr  to
$10^5$~yr.  There is no correlation between $P_3$ and $P_1$.  The
triple systems with $P_1<30^d$ have large period ratios  and are very stable
dynamically (Fig.~\ref{fig:p1p3}).  


\item
The TCs are more massive  than spectroscopic primaries in  a small
fraction (17\% $\pm$ 4\% ) of multiple systems, suggesting that there
is a tendency of the most massive component to be found preferentially
in shortest-period sub-systems.
\end{enumerate}

\section{Discussion}
\label{sec:disc}

It seems that the properties of close SBs are established early in
their evolution.  In a compilation of  pre-Main Sequence (PMS) 
SBs by Melo et al.  (\cite{Melo01}), the period distribution is not
different  in any significant way from the distribution in our
sample. For example, SB periods are divided almost equally between
$1^d - 7^d$ and $7^d - 30^d$ intervals.  The observed frequency of TCs
in the Melo et al. PMS sample, 42\% $\pm$ 19\% (Sterzik el at.
\cite{SMTB05}), is not different from the {\em observed} TC frequency
of 52\% in our sample of SBs in the field.

Our main result (Fig.~\ref{fig:f3p1}) shows that essentially all 
very close SBs are members of higher-order systems.  The KCTF
mechanism is most likely responsible for shortening SB periods during
their life on the Main Sequence.  It acts independently of the TC
period and mass: even a very wide and low-mass companion can influence
the SB, given sufficient time.  The period of Kozai cycles is of the
order of $P_3(P_3/P_1) \sim 10^6$~yr for a typical tertiary period of
$P_3= 1000$~yr and the period of an SB progenitor of $P_1 = 1$~yr. It
takes many ($>10^2$) cycles to complete the KCTF evolution, hence it
could not occur at the PMS stage.  The formation of very close PMS SBs
is not possible  because the PMS stellar radii are too large
and a contact PMS binary would rapidly merge (Whelan \cite{Whelan70}).

For SB periods longer than 3 days, we find an increasing
proportion of pure SBs without tertiaries. The existence of non-triple
SBs is clearly demonstrated by this study. We see that the frequency
of wide TCs to binaries with $P_1>12^d$ is {\em lower} than  the
companion frequency to  single solar-type stars.  These pure SBs have
not passed through KCTF evolution.  What, then, was the mechanism
which produced their orbital decay?

Sterzik, Durisen \& Zinnecker (\cite{SDZ04}) discuss various processes
of binary formation in a systematic way and provide suitable scaling
laws. Molecular cores fragment into mini-clusters of the size of order
100~AU.  The system crossing time turns out to be shorter than the
accretion time scale, hence stars build their masses and interact
dynamically simultaneously.  An accreting multiple system becomes
tighter (Umbreit et al.  \cite{Umbreit}).  Close approaches of the
components of an unstable multiple system lead to ejections and
formation of a close binary with a semi-major axis $\sim 10$ times
shorter than the system size.  Joint action of accretion and
disruption can thus produce close SBs without additional companions.
To form a 10$^d$ SB (semi-major axis 0.1~AU), we need a typical
precursor multiple system of $\sim$1~AU size.

Not  all primordial  multiples  are unstable.   If  orbits of  distant
companions contain  the bulk of a system's angular momentum  (as should
often  be the  case),  there  are no  close  approaches   causing
disintegration. The hardening  of inner orbits in such  systems can be
primarily driven  by accretion  and interactions with  a disk.
An  interesting alternative  has  been   suggested by  Namouni
(\cite{Namouni05}): an orbit can  become eccentric if the system moves
with acceleration, e.g. caused by an  asymmetric jet or a TC, and then
tidal forces  will produce  a close  SB in analogy   to  the KCTF
process.

Some stable TCs will be  lost at later stages by dynamical interaction
with other  members or  systems of a  primordial star  cluster (Kroupa
\cite{Kroupa01}), again leaving pure SBs.

Accretion on a close binary  biases its mass ratio towards 1.  Indeed,
the  mass-ratio distribution  of SBs  is nearly  uniform  (Goldberg et
al. \cite{Goldberg}).   SBs with equal-mass components  (twins) can be
explained   this  way   (Tokovinin  \cite{twins};  Bate \& Bonnell  
\cite{BB97}).    It  is     remarkable  that   the  mass   ratio
distributions   of   SBs  with   and   without   TCs   are  the   same
(Fig.~\ref{fig:p-q}),  pointing  to a  common  process  of their  mass
buildup.  A general  tendency to find the most  massive component of a
multiple system as  a primary in the SB is confirmed  here.  It can be
explained  both by accretion  physics and  by pure  $N$-body dynamics.
However, we find that distant  TCs have often comparable masses or are
even more  massive than the SB  (Fig.~\ref{fig:q3}), contradicting the
simulations of  Delgado-Donate et al.   (\cite{Delgado03}) who predict
{\em only} low-mass TCs.

It becomes thus increasingly clear that close binaries were formed by
a combination of different processes (some probably not yet
identified) and that their orbits and multiplicity were further
modified by subsequent evolutionary effects such as KCTF.  How
efficient is the KCTF mechanism in producing close SBs?  Case-by-case
modeling (Kisseleva et al.  \cite{Kisseleva98}) should now be
complemented by a statistical analysis.  Additional information on the
KCTF will be obtained from the relative orientation of orbits in close
triple systems which are now observable with long-baseline
interferometry (e.g. Muterspraugh et al. \cite{Muterspraugh}).

This work can be considered as a small step towards complete
multiplicity statistics in the solar neighborhood, still largely
unknown because of strong observational selection effects. For
example, is the paucity of $\sim 100^d$ inner periods in multiple
stars (Fig.~\ref{fig:p1p3}) real?  Is it a fossil record of
inner-orbit decay?  A systematic census of {\em all} companions in
nearby solar-type stars using a combination of observing techniques is
required to answer these questions.

\begin{acknowledgements}

We  thank the  staff of  the  Paranal observatory  (and especially  N.
Ageorges) for efficient execution of  our program. The comments of the
Referee, H.~Zinnecker, helped to improve the article.  We are grateful
to  R.~Blum  for  language  editing  of the  manuscript.   The  SIMBAD
database maintained by the University of Strasbourg has been consulted
extensively.  This work makes use of data products from the Two Micron
All Sky Survey and Digital Sky Survey.

\end{acknowledgements}

\appendix

\section{Discussion of detection limits}

  The  knowledge of  the  detection  limits  of various  observing
techniques is essential to  characterize the completeness of the TC
statistics   (Sect.~6)   and   to   correct  the   final   results
(Sect.~7.4). The  limits of each  technique are discussed  and modeled
below.

\subsection{Adaptive optics}

We studied the detection limits  for triple companions in our NACO
observations of spectroscopic binaries. At each radial distance from
the  spatially unresolved SB the rms fluctuation $\sigma$ in the
flux over the corresponding circle was computed, and the detection
limit was assumed to be $5 \sigma$.  This procedure has been checked
by simulating artificial companions and then detecting them visually.
The resulting detection limit is described by a model

\begin{eqnarray}
\Delta K  \leq & 7 r, \;\;  & \rho  \leq  0\farcs1  \nonumber \\
\Delta K  \leq & 5 r +0.9, \;\; & 0\farcs1 <  \rho   < 1\farcs5 \nonumber \\
\Delta K  \leq & 9.05, \;\;   & \rho   \ge 1\farcs5 ,
\label{eq:rect}
\end{eqnarray}
where  $r  =  \log   (\rho/0\farcs035)$.   The  diffraction  limit  at
$\lambda=2.12$~$\mu$m  is $\lambda/D =  0\farcs054$.  A  similar model
was  used by Shatsky  \& Tokovinin  (\cite{ST02}). Our  NACO detection
limits are corroborated by Brandeker (\cite{Brandeker}, his Fig.~4).

\subsection{Visual and CPM companions}

The detectable  magnitude differences  of  known  visual companions
increases  with their  separation $\rho$.  The upper  envelope
corresponds to

\begin{eqnarray}
\Delta V & \leq &
4.5 \log (\rho/0\farcs05).
\label{eq:detvis}
\end{eqnarray}
The components  lying at this  limit were all discovered  with speckle
interferometry (cf.   INT4).  Thus, (\ref{eq:detvis})  may be somewhat
optimistic for 1/2  of our systems that have  never been observed with
speckle techniques according  to INT4.  No new companions  are given in
the   CHARM   catalog   of   lunar   occultaions   and   long-baseline
interferometry  by  Richichi  et  al. (\cite{CHARM}).   An  additional
constraint  that  the apparent  magnitude  of  a  companion should  be
brighter  than  $V=15^m$  is  introduced  because  fainter  stars  are
generally not detected as common-proper-motion (CPM) companions.

\subsection{Radial velocities}

A TC modulates the center-of-mass velocity  of the SB
with a longer  period.  The probability of detecting  such modulation depends
on  the  precision  of  radial-velocity observations  $\sigma$,  their
number $N$ and the time  span $\Delta T$.  Studies of detection limits
with simulations  (Halbwachs et al. \cite{H03}) established that
the relevant  parameter is  the radial velocity  amplitude $A_0$  of a
circular-orbit binary with the same period and $90^\circ$ inclination,

\begin{equation}
A_0 = \frac{213 {\cal M}_3 P_3^{-1/3} }{{\cal M}^{2/3} },
\label{eq:A0}
\end{equation}
where $A_0$ is in km/s, tertiary mass ${\cal M}_3$ and the SB system
mass ${\cal M}$ are in solar masses, and the tertiary's period $P_3$
is in days.  Following Halbwachs et al., we model the spectroscopic
detection limit conservatively as $A_0/\sigma >3$ for all periods
shorter than the data span $\Delta T$, and as $ A_0/\sigma > 3 + 20
\log (P_3/\Delta T)$ for periods shorter than $2 \Delta T$.   The
actual detection limit may extend to periods $P_3$ much longer than
$\Delta T$ if the data are analyzed carefully, but we prefer to be
conservative here by assuming the sharp loss of detection capacity at
$P_3 > 2 \Delta T$.

For  each  SB, we  determined  the  relevant  parameters $\sigma$  and
$\Delta T$  either  from the data  given in SB9  (Pourbaix et
al. \cite{SB9}) or  from the original publications. 
The quality of spectroscopic orbits varies greatly, from crude orbital
solutions  made some 80  years ago  (e.g. HIP  5689, 72524,  86263) to
high-precision  velocities  (HIP  80686)  or long-term  coverage  (HIP
107095).   Moreover, orbits  are  often computed  by combining  radial
velocities of different quality  or with different systematic offsets.
Thus, any  model of  spectroscopic detection limit  is necessarily
very crude and $A_0/\sigma$ is only an indicative parameter, at best.

Despite uncertainties in the spectroscopic  detection limits, we note that
half of our  sample has $\Delta T >5.5$~yr, for 78\%  of SBs $\Delta T
>2$~yr. The median  error  is $\sigma =  1$~km/s. The TCs with
periods  of  a few  years  would  have a good  chance of being  discovered
spectroscopically,   hence the paucity of such  companions in our
sample must be real.

\subsection{Astrometry}

Astrometry is a poweful technique to detect low-mass companions
with orbital periods over several years.  Makarov \& Kaplan
(\cite{MK05}) recently published a catalog of astrometric binaries
revealed either by a significant difference between ``instantaneous''
proper motion (PM) measured by Hipparcos and long-term PM in the
Tycho-2 catalog ($\Delta \mu$ binaries) or by the acceleration
measured directly by Hipparcos ($\dot{\mu}$ binaries, also known as
G-solutions in Hipparcos). There are 11 astrometric binaries in common
with our sample, of which six are already listed here with TCs
discovered by other techniques (including HIP 19248 and 64219 resolved
for the first time here).  We did not resolve two $\dot{\mu}$
binaries, HIP 35487 and 114639, with NACO, possibly because their
astrometric companions are white dwarfs (cf. discussion in Notes to
Table~5).

Gontcharov  et  al.   (\cite{G01})  published a  similar  study:  they
brought historical astrometric catalogs  into the Hipparcos system and
identified several stars as astrometric binaries.  However, only a few
astrometric orbits of these stars  are published to date (none for our
sample),  and Makarov  \& Kaplan  (\cite{MK05}) did  not independently
confirm the PM variations of many stars. We observed with NACO but did
not  resolve HIP  67153, 80686,  85365,  86263 listed  as binaries  by
Gontcharov  et al.   Here we  consider astrometric  detections without
published orbits only  as hints of the existence of  TCs.  Hence we do
not take into account astrometric companions in our statistics.

\section{Maximum Likelihood method}

{\em Problem  layout.} Suppose  that a total  of $N$ targets  have been
surveyed.  We determine the distribution of companions over parameters
by  dividing the  parameter  space  into $K$  bins,  the fractions  of
companions in  each bin  are denoted as  ${\bf f}  = \{ f_k  \}$.  Let
$N_{i,k}$ be  the number of  detected companions for $i$-th  star and
$k$-th bin (1 when a companion is detected, 0 otherwise). The
total number of  companions per bin is $N_k =  \sum_i N_{ik}$. Thus, a
raw estimate of companion frequency will be

\begin{equation}
f_{k, \rm raw} = N_k /N.
\label{eq:fraw}
\end{equation}

In fact  the probabilities  of companion detection  for each  star and
each bin  $d_{i,k}$ are generally less  than 1, so we  have to account
for incomplete detections. This is best done by the Maximum Likelihood
(ML) technique.  First, we evaluate the  probabilities of observations
for each star $p_i$ and compute the {\em likelihood function}

\begin{equation}
{\cal L} = \prod_i p_i ({\bf f}).
\label{eq:L}
\end{equation}
By  minimizing the  natural logarithm  $S =  - 2  \ln {\cal  L}$ over
parameters ${\bf f}$ we estimate  the parameters.  The contours of $S$
in  the ${\bf  f}$-space  define confidence  limits  of the  parameter
estimates:  $\Delta S  = 1$  corresponds  to the  68\% interval  (``$1
\sigma$''), $\Delta S =  2.71$ to 90\% and $\Delta S =  4$ to 95\%, in
direct  analogy with  the Gaussian  probability distribution  (Avni et
al. \cite{Avni}).

\begin{figure}[ht]
\centerline{\psfig{figure=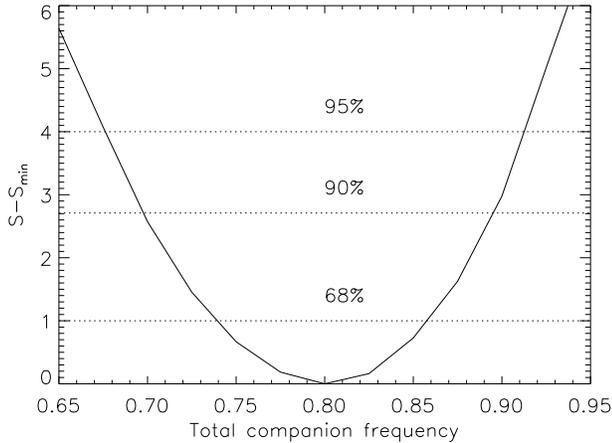,width=8.5cm} }
\caption{The shape of the $S - S_{min}$ function near its minimum and the
  confidence intervals on the total companion frequency for a
  sub-sample of close SBs, $P_1 <7$d.
\label{fig:S} }
\end{figure}

{\em Independent detections.} In the simplest case we assume companion
detections in each bin to be independent of other bins. The average
expected number of detected companions is $f_k d_{i,k}$, and the
probabilities are computed from the Poisson distribution,

\begin{equation}
p_i  =  \frac{ (f_k d_{i,k})^{N_{i,k}} }{ N_{i,k}! }  \exp \; ( - f_k d_{i,k} ).
\label{eq:Poi}
\end{equation}
Considering that $N_{i,k}$ only takes values of 0 or 1, the likelihood
equation can be written as

\begin{equation}
- 2\ln {\cal L} = -2 \sum_{i=1}^N \sum_{k=1}^K [ N_{i,k} \ln (f_k
 d_{i,k}) - f_k d_{i,k} ] + {\rm const} .
\label{eq:Lind}
\end{equation}
By differentiating (\ref{eq:Lind}) over $f_k$ we obtain the estimates
\begin{equation}
\hat{f}_k = \frac{N_k} { \sum_i d_{i,k} }.
\label{eq:fk}
\end{equation}
This  result is intuitive:  we simply  correct the  observed companion
frequencies $f_{\rm  raw}$ by  the average probabilities  of detection
$\sum_i d_{i,k}/N$.

We compute the probabilities $d_{i,k}$ as fractional surfaces of each
bin above the detection limit curve $q_{\rm lim} (P_3)$, individually
for each star.  The underlying assumption is that the (unknown)
companion distribution within each bin is uniform.  Looking at
Fig.~\ref{fig:q3}, we note that there are no strong gradients of
companion frequency over $q_3$ or $P_3$, hence the assumption is good.

{\em Mutually exclusive detections.}  A given spectroscopic binary can
have  no more  than one  tertiary  companion, by  definition. Thus,  a
detection of  a companion in some bin  automatically excludes companions
in other bins. Denoting by  $f_0$ the probability of no companions, we
have an additional condition $\sum_k f_k + f_0 = 1$.

According  to   the  Bayes  theorem,  the   probability  of  obtaining
an observational result  for the $i$-th star  is related to  the distribution
$f_k$ and  the conditional  probabilities (detecting or  not detecting
companion) $p(i | k)$,

\begin{equation}
p_i = \sum_{k=0} p(i | k) f_k .
\label{eq:bayes}
\end{equation}

In the case of companion non-detection, the probability $p^{-}_i$ is
\begin{equation}
p^{-}_i = f_0 + \sum_{k=1} (1 - d_{i,k}) f_k = 1 -  \sum_{k=1} d_{i,k} f_k,
\label{eq:p-}
\end{equation}
while for stars with  companions the probabilities are simply $p^{+}_i
= f_k d_{i,k}$, according to (\ref{eq:Poi}). This leads to the
likelihood function

\begin{eqnarray}
-2 \ln {\cal L} & = & -2 \sum_{i^-} \ln (1 -  \sum_{k} d_{i,k} f_k) \nonumber \\
    & &  - 2  \sum_{i} \sum_k  N_{i,k} \ln (d_{i,k} f_k ) + {\rm const.} ,
\label{eq:L2}
\end{eqnarray}
where  $i^-$  means that  the  summation  extends  over stars  without
detected  companions. The  second term  reduces to  a single  sum over
detected companions  because otherwise $N_{i,k} =  0$.  The parameters
$f_k$  are estimated  by direct  minimization of  (\ref{eq:L2}). These
ML estimates are very close to those given by  eq.~\ref{eq:fk}.

{\em Confidence  intervals.} In order to find  confidence intervals on
$f_k$ or on the total companion frequency $f = \sum_k f_k$, we have to
perform constrained minimization of (\ref{eq:L2}) with a fixed parameter
$f_k^*$.  The curve  $S(f_k^*) -  S_{\rm min}$  then  defines the
confidence intervals, as illustrated in Fig.~\ref{fig:S}.


\end{document}